\documentclass[%
 reprint,
superscriptaddress,
nofootinbib,
 amsmath,amssymb,
 aps,
 showkeys,
]{revtex4-2}

\usepackage{graphicx}
\usepackage{dcolumn}
\usepackage{bm}
\usepackage{hyperref}
\usepackage[bottom]{footmisc}
\usepackage{footnote}

\newcommand{\be}{\begin{equation}}
\newcommand{\ee}{\end{equation}}
\newcommand{\bea}{\begin{eqnarray}}
\newcommand{\eea}{\end{eqnarray}}
\newcommand{\dAu}{d$+$Au~}
\newcommand{\jpsi}{$J/\psi~$}

\begin{document}

\title{Probing the gluonic structure of the deuteron with \jpsi photoproduction\\ in d$+$Au ultra-peripheral collisions}
\affiliation{Abilene Christian University, Abilene, Texas   79699}
\affiliation{AGH University of Science and Technology, FPACS, Cracow 30-059, Poland}
\affiliation{Alikhanov Institute for Theoretical and Experimental Physics NRC "Kurchatov Institute", Moscow 117218, Russia}
\affiliation{Argonne National Laboratory, Argonne, Illinois 60439}
\affiliation{American University of Cairo, New Cairo 11835, New Cairo, Egypt}
\affiliation{Ball State University, Muncie, Indiana, 47306}
\affiliation{Brookhaven National Laboratory, Upton, New York 11973}
\affiliation{University of Calabria \& INFN-Cosenza, Italy}
\affiliation{University of California, Berkeley, California 94720}
\affiliation{University of California, Davis, California 95616}
\affiliation{University of California, Los Angeles, California 90095}
\affiliation{University of California, Riverside, California 92521}
\affiliation{Central China Normal University, Wuhan, Hubei 430079 }
\affiliation{University of Illinois at Chicago, Chicago, Illinois 60607}
\affiliation{Creighton University, Omaha, Nebraska 68178}
\affiliation{Czech Technical University in Prague, FNSPE, Prague 115 19, Czech Republic}
\affiliation{Technische Universit\"at Darmstadt, Darmstadt 64289, Germany}
\affiliation{ELTE E\"otv\"os Lor\'and University, Budapest, Hungary H-1117}
\affiliation{Frankfurt Institute for Advanced Studies FIAS, Frankfurt 60438, Germany}
\affiliation{Fudan University, Shanghai, 200433 }
\affiliation{University of Heidelberg, Heidelberg 69120, Germany }
\affiliation{University of Houston, Houston, Texas 77204}
\affiliation{Huzhou University, Huzhou, Zhejiang  313000}
\affiliation{Indian Institute of Science Education and Research (IISER), Berhampur 760010 , India}
\affiliation{Indian Institute of Science Education and Research (IISER) Tirupati, Tirupati 517507, India}
\affiliation{Indian Institute Technology, Patna, Bihar 801106, India}
\affiliation{Indiana University, Bloomington, Indiana 47408}
\affiliation{Institute of Modern Physics, Chinese Academy of Sciences, Lanzhou, Gansu 730000 }
\affiliation{University of Jammu, Jammu 180001, India}
\affiliation{Joint Institute for Nuclear Research, Dubna 141 980, Russia}
\affiliation{Kent State University, Kent, Ohio 44242}
\affiliation{University of Kentucky, Lexington, Kentucky 40506-0055}
\affiliation{Lawrence Berkeley National Laboratory, Berkeley, California 94720}
\affiliation{Lehigh University, Bethlehem, Pennsylvania 18015}
\affiliation{Max-Planck-Institut f\"ur Physik, Munich 80805, Germany}
\affiliation{Michigan State University, East Lansing, Michigan 48824}
\affiliation{National Research Nuclear University MEPhI, Moscow 115409, Russia}
\affiliation{National Institute of Science Education and Research, HBNI, Jatni 752050, India}
\affiliation{National Cheng Kung University, Tainan 70101 }
\affiliation{Nuclear Physics Institute of the CAS, Rez 250 68, Czech Republic}
\affiliation{Ohio State University, Columbus, Ohio 43210}
\affiliation{Institute of Nuclear Physics PAN, Cracow 31-342, Poland}
\affiliation{Panjab University, Chandigarh 160014, India}
\affiliation{NRC "Kurchatov Institute", Institute of High Energy Physics, Protvino 142281, Russia}
\affiliation{Purdue University, West Lafayette, Indiana 47907}
\affiliation{Rice University, Houston, Texas 77251}
\affiliation{Rutgers University, Piscataway, New Jersey 08854}
\affiliation{Universidade de S\~ao Paulo, S\~ao Paulo, Brazil 05314-970}
\affiliation{University of Science and Technology of China, Hefei, Anhui 230026}
\affiliation{South China Normal University, Guangzhou, Guangdong 510631}
\affiliation{Shandong University, Qingdao, Shandong 266237}
\affiliation{Shanghai Institute of Applied Physics, Chinese Academy of Sciences, Shanghai 201800}
\affiliation{Southern Connecticut State University, New Haven, Connecticut 06515}
\affiliation{State University of New York, Stony Brook, New York 11794}
\affiliation{Instituto de Alta Investigaci\'on, Universidad de Tarapac\'a, Arica 1000000, Chile}
\affiliation{Temple University, Philadelphia, Pennsylvania 19122}
\affiliation{Texas A\&M University, College Station, Texas 77843}
\affiliation{University of Texas, Austin, Texas 78712}
\affiliation{Tsinghua University, Beijing 100084}
\affiliation{University of Tsukuba, Tsukuba, Ibaraki 305-8571, Japan}
\affiliation{United States Naval Academy, Annapolis, Maryland 21402}
\affiliation{Valparaiso University, Valparaiso, Indiana 46383}
\affiliation{Variable Energy Cyclotron Centre, Kolkata 700064, India}
\affiliation{Warsaw University of Technology, Warsaw 00-661, Poland}
\affiliation{Wayne State University, Detroit, Michigan 48201}
\affiliation{Yale University, New Haven, Connecticut 06520}

\author{M.~S.~Abdallah}\affiliation{American University of Cairo, New Cairo 11835, New Cairo, Egypt}
\author{B.~E.~Aboona}\affiliation{Texas A\&M University, College Station, Texas 77843}
\author{J.~Adam}\affiliation{Brookhaven National Laboratory, Upton, New York 11973}
\author{L.~Adamczyk}\affiliation{AGH University of Science and Technology, FPACS, Cracow 30-059, Poland}
\author{J.~R.~Adams}\affiliation{Ohio State University, Columbus, Ohio 43210}
\author{J.~K.~Adkins}\affiliation{University of Kentucky, Lexington, Kentucky 40506-0055}
\author{G.~Agakishiev}\affiliation{Joint Institute for Nuclear Research, Dubna 141 980, Russia}
\author{I.~Aggarwal}\affiliation{Panjab University, Chandigarh 160014, India}
\author{M.~M.~Aggarwal}\affiliation{Panjab University, Chandigarh 160014, India}
\author{Z.~Ahammed}\affiliation{Variable Energy Cyclotron Centre, Kolkata 700064, India}
% ...

\collaboration{STAR Collaboration}\noaffiliation 
\date{\today}
\begin{abstract}
Understanding gluon density distributions and how they are modified in nuclei are among the most important goals in nuclear physics. In recent years, diffractive vector meson production measured in ultra-peripheral collisions (UPCs) at heavy-ion colliders has provided a new tool for probing the gluon density. In this Letter, we report the first measurement of  \jpsi photoproduction off the deuteron in UPCs at the center-of-mass energy $\sqrt{s_{_{\rm NN}}}=200~\rm GeV$ in d$+$Au collisions. The differential cross section as a function of momentum transfer $-t$ is measured. In addition, data with a neutron tagged in the deuteron-going Zero-Degree Calorimeter is investigated for the first time, which is found to be consistent with the expectation of incoherent diffractive scattering at low momentum transfer. Theoretical predictions based on the Color Glass Condensate saturation model and the Leading Twist Approximation nuclear shadowing model are compared with the data quantitatively. A better agreement with the saturation model has been observed. With the current measurement, the results are found to be directly sensitive to the gluon density distribution of the deuteron and the deuteron breakup process, which provides insights into the nuclear gluonic structure.

\end{abstract}

\maketitle

One of the most outstanding problems in modern nuclear physics is the partonic structure of nucleons (protons and neutrons) and nuclei. Specially, the origin of modified partonic structure of nucleons bounded in nuclei has been of extreme interest, with its first discovery on the valance quarks by the European Muon Collaboration (EMC) almost 40 years ago, known as the EMC effect~\cite{Aubert:1983xm,Ashman:1988bf,Gomez:1993ri,Arneodo:1988aa,Arneodo:1989sy,Allasia:1990nt,Seely:2009gt}. However, this modification was not only found in valance quarks but also in gluons~\cite{Ethier:2020way}, where gluons start to dominate in parton densities at high energies~\cite{Abramowicz:2015mha} and become more relevant in considering the parton hard scattering processes. See Ref.~\cite{Accardi:2012qut} for a review. 

Coherent diffractive Vector-Meson (VM) production off nuclei has been considered as one of the golden
measurements to study the gluon density and its spatial distributions~\cite{Accardi:2012qut, PhysRevLett.36.1233,PhysRevLett.48.73,Arneodo:1994qb,Adler:2002sc,Timoshenko:2005si,Khachatryan:2016qhq,Abelev:2012ba,ALICE:2015nbw,Adamczyk:2017vfu,Acharya:2020sbc,Acharya:2021gpx,Acharya:2021bnz,Acharya:2021ugn,LHCb:2021hoq}.
In recent analyses carried out by the Large Hadron Collider (LHC) collaborations~\cite{Khachatryan:2016qhq,Abelev:2012ba,Adamczyk:2017vfu,Acharya:2020sbc,Acharya:2021gpx,Acharya:2021bnz,Acharya:2021ugn,LHCb:2021hoq}, photoproduction of the \jpsi meson has been measured in ultra-peripheral collisions (UPCs) of heavy ions -  a photon-ion interaction at large impact parameter arising from extreme electromagnetic fields~\cite{Bertulani:2005ru}. The resulting cross sections were found to be significantly suppressed with respect to that of a free proton~\cite{Khachatryan:2016qhq,Abelev:2012ba,Acharya:2021bnz,Acharya:2021ugn}. Leading Twist Approximation (LTA) calculations strongly suggest that the suppression is caused by the nuclear shadowing effect~\cite{Strikman:2018mbu,Guzey:2013qza,Guzey:2018tlk}, while other models, e.g., the Color Dipole Model with gluon saturation and nucleon shape fluctuations~\cite{Sambasivam:2019gdd}, can also describe the UPC data qualitatively. As of today, neither the gluonic structure of heavy nuclei nor the modification of their partonic structure is fully understood. 

An interesting experimental approach to reveal the gluonic structure of nuclei is to study the deuteron - the simplest nuclear bound state of one proton and one neutron. While neither gluon saturation nor the nuclear shadowing effect is expected to be significant in such a loosely bound system, the deuteron may provide unique physics insights to phenomena that are poorly understood from data of heavy nuclei, e.g., the interplay between coherent and incoherent VM production, nuclear breakup, single and double nucleon scattering, and short-range nuclear correlations. For example, recent studies have shown potential connections between (gluon) EMC effects and short-range nuclear correlations in light nuclei~\cite{Tu:2020ymk,Cosyn:2020kwu,Strikman:2017koc}. This is a subject of interest for a wide range of physics communities, from nuclear and particle physics to high-density neutron stars in astrophysics.

 \begin{figure}[thb]
\includegraphics[width=3in]{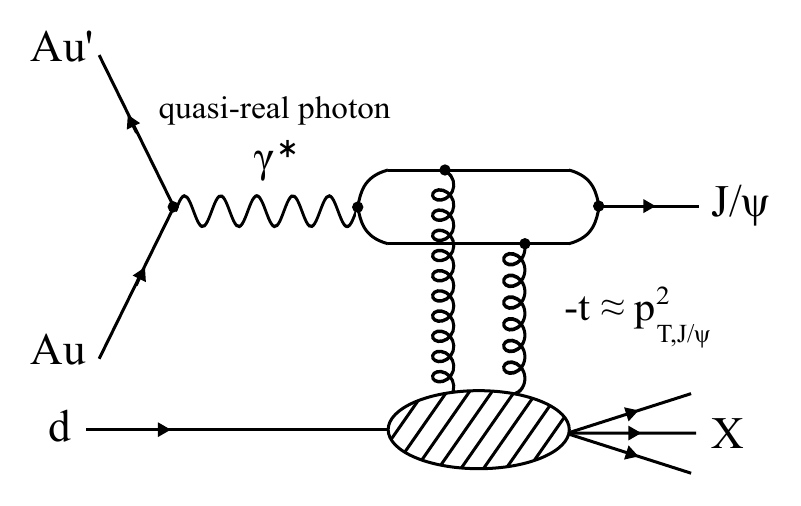}
  \caption{ \label{fig:figure_1}  Photoproduction of \jpsi in d$+$Au UPCs, where $X$ represents the deuteron (coherent) or deuteron-dissociative (incoherent) system. }
\end{figure}

In this Letter, we investigate the differential cross section of \jpsi photoproduction as a function of momentum transfer, $-t$, in \dAu UPC events at $\sqrt{s_{_{\rm NN}}}=200\rm~GeV$. The \jpsi photoproduction process in d$+$Au UPCs is illustrated in Fig.~\ref{fig:figure_1}. In the photoproduction limit, the momentum transfer variable $-t$ can be approximated by the transverse momentum squared of $J/\psi$ particles, $p^{2}_{\rm T,J/\psi}$. The approximate photon-nucleon center-of-mass energy is~\cite{TheALICE:2014dwa}, $W=\sqrt{2\langle E_{N}\rangle M_{J/\psi}e^{-y}} \sim 25~\rm{GeV}$, where $E_{N}$ is the average beam energy per nucleon, $M_{J/\psi}$ is the mass of the $J/\psi$ particle, and $y$ is the $J/\psi$ rapidity. In addition, the differential $J/\psi$ cross section with single neutron tagged events is reported. The data are compared with two theoretical models: i) Color Glass Condense (CGC) saturation model and ii) LTA nuclear shadowing model. These model predictions are based on an extension from heavy nuclei to light nuclei~\cite{Mantysaari:2019jhh,Guzey:2021star,Guzey:2018tlk}.\footnote{Both model calculations are made specifically to the \dAu UPC data at RHIC, where Ref.~\cite{Guzey:2021star} is an extension of Ref.~\cite{Guzey:2018tlk} from heavy nuclei at the LHC to the deuteron at RHIC.}

The Solenoidal Tracker At RHIC (STAR) detector~\cite{Ackermann:2002ad} and its subsystems have been thoroughly described in previous STAR papers~\cite{Adam:2018tdm,Adam:2020cwy}. This analysis utilizes several subsystems of the STAR detector. Charged particle tracking, including transverse momentum reconstruction and charge sign determination, is provided by the Time Projection Chamber (TPC)~\cite{Anderson:2003ur} positioned in a 0.5 Tesla longitudinal magnetic field. The TPC volume extends from 50 to 200 cm from the beam axis and covers pseudorapidities $|\eta|<1.0$ and over the full azimuthal angle, $0<\phi<2\pi$. Surrounding the TPC is the Barrel Electromagnetic Calorimeter (BEMC)~\cite{Beddo:2002zx}, which is a
lead-scintillator sampling calorimeter. The BEMC is segmented into 4800 optically isolated
towers covering the full azimuthal angle for pseudorapidities $|\eta|<1.0$. There are two Beam-Beam Counters (BBCs)~\cite{Whitten:2008zz}, one on each side of the STAR main detector, covering a pseudorapidity range of $3.4<|\eta|<5.0$. There are also two Zero Degree Calorimeters (ZDCs)~\cite{Ackermann:2002ad}, used to determine and monitor the
luminosity and tag the forward neutrons.

\begin{figure}[hb]
\includegraphics[width=1.67in]{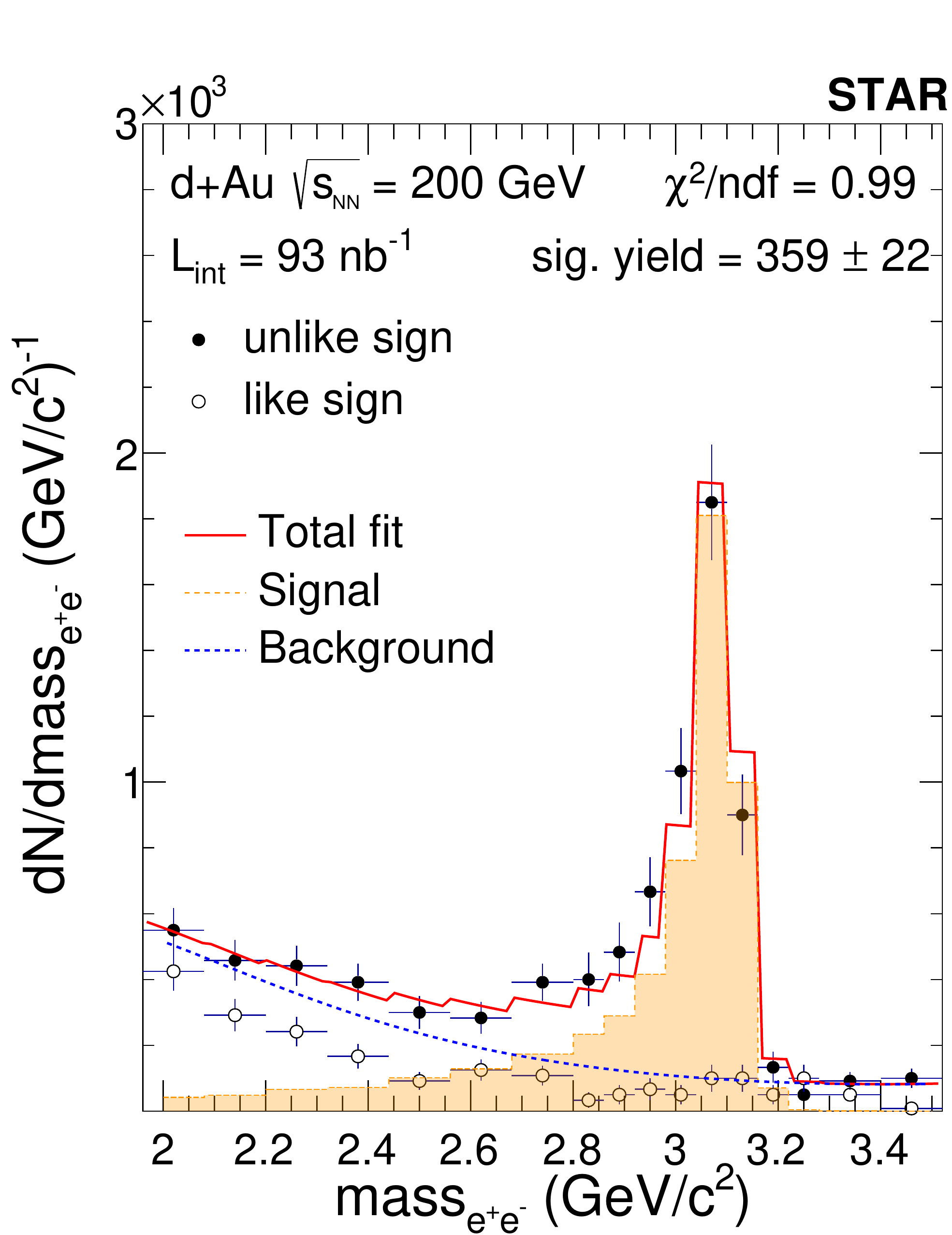}
\includegraphics[width=1.67in]{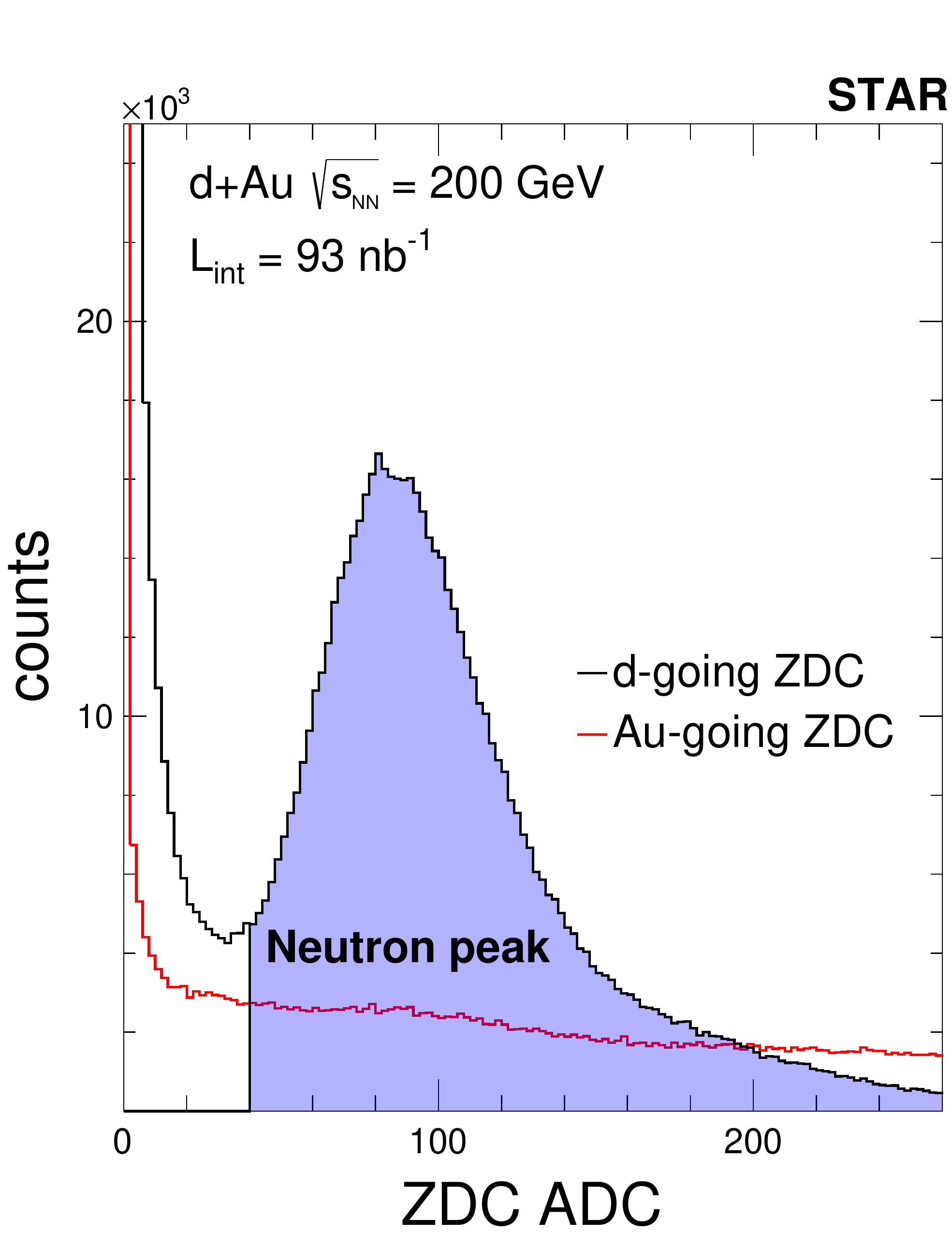}
  \caption{ \label{fig:figure_2} Left: invariant mass distribution and corresponding fits of \jpsi candidates reconstructed via the electron decays. Right: Zero-Degree Calorimeter (ZDC)  energy deposition (arbitrary units) distribution
for both Au- and deuteron-going directions.  }
\end{figure}

The UPC data were collected by the STAR experiment during the 2016 \dAu run, corresponding to an integrated luminosity of 93 $\rm nb^{-1}$ and approximately $2\times10^6$ UPC \jpsi-triggered events. The \jpsi candidates are reconstructed via the electron decay channel, \jpsi$\rightarrow e^{+}e^{-}$, which has a branching ratio of 5.93\%\cite{10.1093/ptep/ptaa104}. Based on this channel, the UPC \jpsi trigger is defined by no signal in either BBC East or West, Time-Of-Flight (TOF)~\cite{Ackermann:2002ad} track multiplicity between 2 and 6, and a topological selection of back-to-back clusters in the BEMC. In the offline analysis, the events are required to have a valid vertex that is reconstructed within 100 cm of the center of the STAR detector. In addition, a valid event is required to have at least two TPC tracks associated with the primary vertex with transverse momentum $p_{\rm T}>0.5~{\rm GeV}/c$ and $|\eta|<1.0$. Single electron candidates are selected from charged tracks reconstructed in the TPC, which are required to have at least 25 space points (out of a maximum of 45) to ensure sufficient momentum resolution, contain no fewer than 15 points for the ionization energy loss ($dE/dx$) determination to ensure good $dE/dx$ resolution, and to be matched to a BEMC cluster.
Furthermore, these tracks are required to have a distance of closest approach less than 3 cm from the primary vertex. To further enhance the purity of electron candidates for the \jpsi reconstructions, an unlike-sign electron pair selection is performed based on the $dE/dx$ of charged tracks. The variable $n_{\sigma,e}$ ($n_{\sigma,\pi}$) is the difference between the measured $dE/dx$ value compared to an electron ($\pi$) hypothesis of the predicted $dE/dx$ value. It is calculated in terms of number of standard deviations from the predicted mean. The pair selection variable $\chi^{2}_{e^{+}e^{-}}$ is defined as $n^{2}_{\sigma,e^{+}}+n^{2}_{\sigma,e^{-}}$ (similar for $\pi$). For the region of $\chi^{2}_{\pi^{+}\pi^{-}} < 30$, the ratio $\chi^{2}_{e^{+}e^{-}}/\chi^{2}_{\pi^{+}\pi^{-}}$ is required to be less than 1/3, while for $\chi^{2}_{\pi^{+}\pi^{-}} > 30$, $\chi^{2}_{e^{+}e^{-}}$ must be less than 10. This pair selection ensures the purity of electrons is higher than 95\%, which is determined by a data-driven approach using photonic electrons~\cite{Adam:2018tdm}.

The unlike-sign electron candidates are paired to reconstruct an invariant mass distribution of \jpsi candidates, while the like-sign pairs are also investigated to indicate the contribution from the combinatorial background. The resulting \jpsi candidates are required to have a rapidity $|y|<1.0$. In Fig.~\ref{fig:figure_2} (left), the invariant mass distribution is shown with a template fit to extract the raw yield of \jpsi particles. The signal template is taken from the STARlight~\cite{Klein:2016yzr} Monte Carlo program that was run through the STAR detector GEANT3 simulation~\cite{Brun:1987ma} for its detector response, indicated by the shaded histogram. Motivated by similar studies in Refs.~\cite{STAR:2019wlg,Abelev:2012ba,ALICE:2013wjo}, the background function is taken to be of the form, $(m-A)e^{B(m-A)(m-C)+Cm^{3}}$, which can describe both the combinatorial and the two-photon interaction ($\gamma \gamma \rightarrow e^{+}e^{-}$) backgrounds. The fitted result is shown as the dotted line, where $m_{e^{+}e^{-}}$ is the invariant mass of two oppositely charged electrons, and $A$, $B$, and $C$ are free parameters~\cite{TheALICE:2014dwa}. The raw yield of the entire analyzed sample after full event selections and background subtraction is $359\pm22$. For measurement of the differential cross section, raw yields of each $p^{2}_{\rm T,J/\psi}$ interval are determined based on the same fitting procedure.  In Fig.~\ref{fig:figure_2} (right), the ZDC energy depositions in terms of Analog-to-Digital Converter (ADC) count are shown for both Au- and deuteron-going directions. For the deuteron-going direction, an ADC count larger than 40 is required for events associated with single neutron emission. Note that after extracting the $J/\psi$ signal, no significant background (pedestal) has been found under the neutron peak for the ADC count larger than 40. 

The differential cross section of \jpsi photoproduction as a function of $-t$ is measured in the d$+$Au UPCs, which can be related to the photon-deuteron cross section based on the following relation, 
\be
\frac{d^{2}\sigma^{(\rm{d+Au\rightarrow J/\psi + X})}}{dtdy} = \Phi_{T,\gamma} \frac{d^{2}\sigma^{(\rm{\gamma^{*}+d\rightarrow J/\psi + X)}}}{dtdy},
\ee
\noindent where $ \Phi_{T,\gamma}$ is the average transversely polarized photon flux emitted from the Au nucleus\footnote{The probability of a photon emitted by the deuteron is $\sim~$4 orders of magnitude smaller, therefore negligible in this analysis.} with \jpsi rapidity $|y|<1.0$, and $X$ represents the deuteron (coherent) or the deuteron-dissociative (incoherent) system. Therefore, the full differential cross section in the photon-deuteron system can be written as, 
\begin{align}
	\frac{d^{2}\sigma^{(\rm{\gamma^{*}+d\rightarrow J/\psi + X)}}}{dtdy} \nonumber 
	& = \frac{1}{\Phi_{T,\gamma}}\frac{N_{obs}}{\Delta t \times \Delta y \times (A\times\epsilon) \times \epsilon_{trig}}  \nonumber \\
	&\times  \frac{1}{L_{int} \times BR(e^{+}e^{-})}.
\end{align}
\noindent Here $\Phi_{T,\gamma}=11.78$ is based on the STARlight MC generator, where the photon flux is calculated based on the Au nucleus thickness function and the photon number density determined from the Weizsacker-Williams method~\cite{Klein:2016yzr}. The $N_{obs}$ is the raw \jpsi yield, $L_{int}$ is the integrated luminosity, $BR(e^{+}e^{-})=5.93\%$ is the branching ratio of \jpsi decaying into an electron pair, $\Delta t$ is the bin width of $p^{2}_{\rm T,J/\psi}$, $\Delta y = 2.0$ is the rapidity range, $A\times\epsilon$ is the \jpsi reconstruction acceptance and efficiency corrections, and $\epsilon_{trig}$ is the trigger efficiency correction. The \jpsi reconstruction efficiency and trigger efficiency corrections are based on the STARlight MC events embedded into STAR zero-bias events, where an unfolding technique is employed in the correction procedure. The default unfolding algorithm is based on the Bayesian method from the RooUnfold software package~\cite{Prosper:2011zz}. 

\begin{figure}[thb]
\includegraphics[width=3.3in]{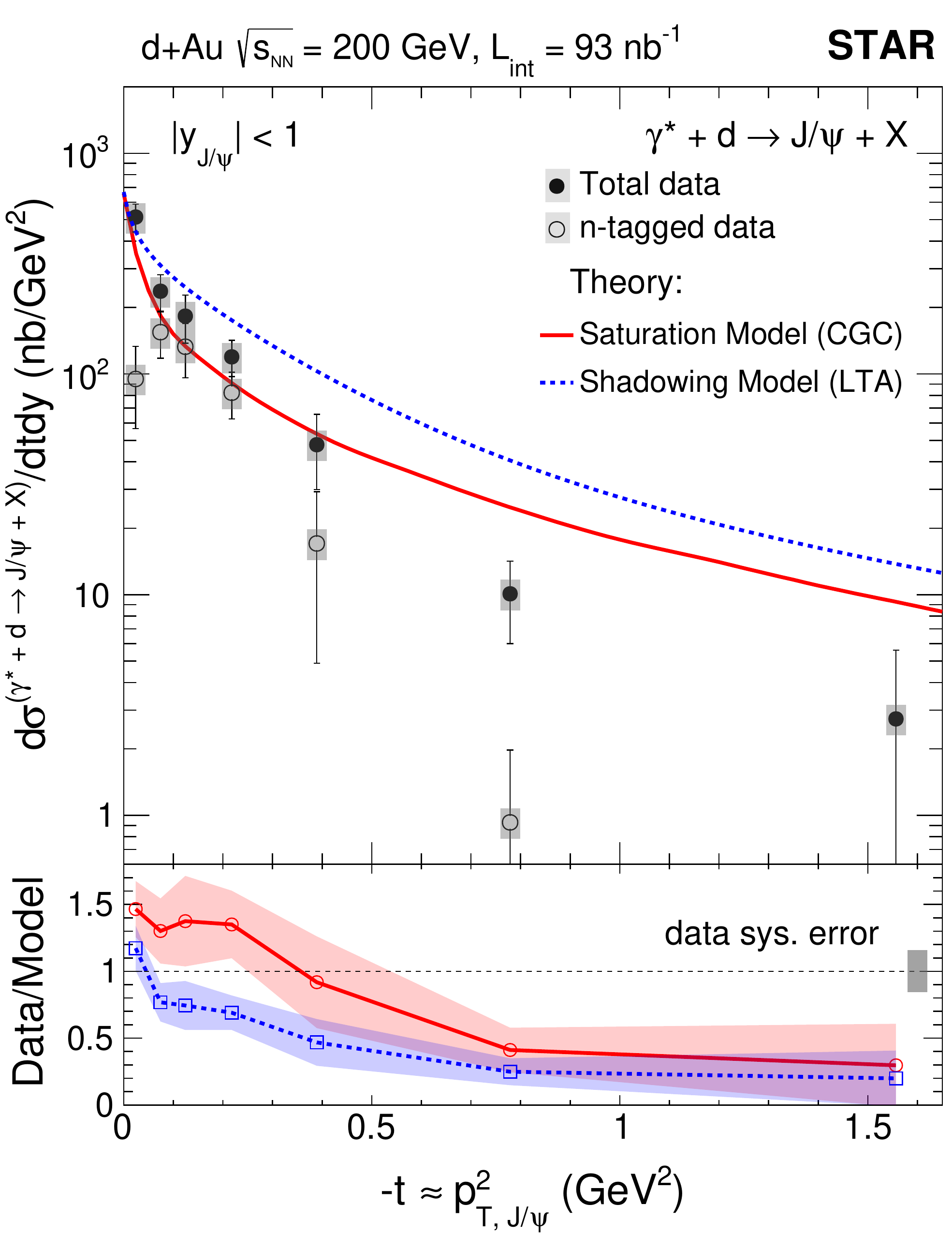}
  \caption{ \label{fig:figure_3} Upper: differential cross section as a function of $p^{2}_{T, J/\psi}$ of \jpsi photoproduction in UPCs at $\sqrt{s_{_{\rm NN}}}=200\rm~GeV$. Data for the total diffractive process are shown with solid markers, while data with neutron tagging in the deuteron-going ZDC are shown with open markers. Theoretical predictions based on the saturation model (CGC)~\cite{Mantysaari:2019jhh} and the nuclear shadowing model (LTA)~\cite{Guzey:2021star} are compared with data, shown as lines. Statistical uncertainty is represented by the error bars, and the systematic uncertainty is denoted by the shaded box. Lower: ratios of total data and models are presented as a function of $-t \approx p^{2}_{T, J/\psi}$. Color bands are statistical uncertainty based on the data only, while systematic uncertainty is indicated by the gray box.  }
\end{figure}

Different sources of systematic uncertainty on the differential cross section were investigated, which were quantitatively motivated by previous STAR publications on VM and di-lepton measurements~\cite{STAR:2019yox,Adamczyk:2017vfu,Adam:2018tdm}. Variations of the fit functions, signal templates, yield extraction methods (bin counting vs fit parameter), and momentum resolution of tracks yield a combined systematic uncertainty of 7.3\%. Track selections with more than 20 or 30 space points in TPC hits, with more than 10 or 20 space points of $dE/dx$ determination, and less than 2 cm in a distance of closest approach with respect to the primary vertex were investigated and found to lead to a systematic uncertainty of 4\%. Variation of the electron identification selection creiteria yields a systematic uncertainty of 2\%. The systematic uncertainty associated with the unfolding technique, e.g., regularization parameter (4 vs 10 iterations), unfolding algorithm (RooUnfold Bayesian vs TUnfold~\cite{Schmitt:2012kp}), and modified underlying truth distributions (exponential vs flat), is found to be 3\%. The trigger efficiency associated with the trigger simulation of the BEMC is found to have an uncertainty of 8\%. The systematic uncertainty on the integrated luminosity determined by the STAR experiment during this \dAu run is 10\%~\cite{STAR:2006kxj,STAR:2003pjh}. Finally, the systematic uncertainty on modeling the transversely polarized photon flux is found to be 2\% by varying the Au radius by $\pm 0.5~\rm{fm}$, where a similar study has been done in Ref.~\cite{TheALICE:2014dwa} at the LHC. The different sources of uncertainty are added in quadrature for the total systematic uncertainty, which is found to be 15.8\%. The systematic uncertainty is largely independent of $-t$, which is expected given that the daughter electrons in the studied kinematic region are within a range of momentum with good detector resolutions.

In Fig.~\ref{fig:figure_3}, the fully corrected differential cross section of \jpsi photoproduction in \dAu UPCs at $\sqrt{s_{_{\rm NN}}}=200\rm~GeV$ is shown. The total diffractive $J/\psi$ cross section is labelled ``Total data". Figure~\ref{fig:figure_3} also shows the $n-$tagged data, which requires that a neutron be detected in the deuteron-going ZDC from deuteron breakup. 
There are three distinct physics processes that contribute to the ``Total data": i) coherent diffraction, $X=\rm{deuteron}$; ii) incoherent diffraction with elastic nucleon, $X=\rm{proton+neutron}$; iii) incoherent diffraction with nucleon dissociation, $X=\rm{proton~(neutron)+fragments}$. For i), it is possible that the deuteron can be broken up by a secondary soft photon, although with small probability, on the order of 0.1\% estimated in the measured kinematic region.~\cite{Klein:2003bz,Baltz:2002pp}.



Although separating the three physics processes experimentally is difficult, the STAR ZDC with approximately $\pm$2.5--3 mrad of angular acceptance~\cite{Adams:2004dv} can capture almost 100\% of the neutron spectators. For the case when the neutron is the leading nucleon, the acceptance is nearly 100\% for $p^{2}_{\rm T,J/\psi} \approx p^{2}_{\rm T,neutron} < 0.1 \rm~GeV^{2}$. 
Therefore, in the very low $-t$ region, the $n$-tagged events are expected to be dominated by the incoherent scattering process~\cite{Klein:2003bz,Baltz:2002pp}. In addition, there is the possibility of more complicated incoherent scattering processes, e.g., the photon interacts with both nucleons simultaneously~\cite{RevModPhys.50.261,RevModPhys.51.407,Rogers:2005bt}, where the data with neutron tagging reported in this measurement will be extremely helpful in constraining these scenarios.  

 \begin{figure*}[thb]
\includegraphics[width=6.in]{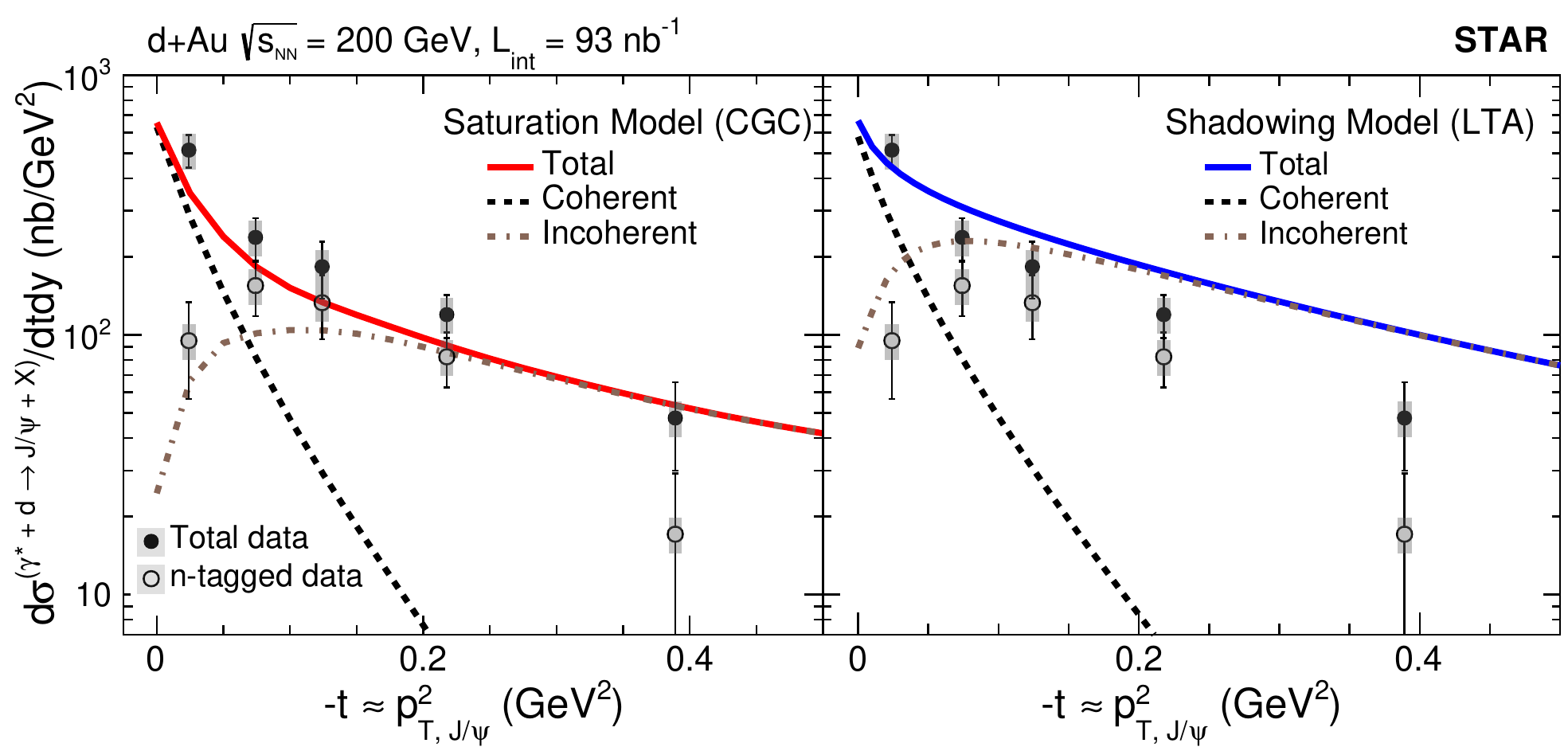}
  \caption{ \label{fig:figure_4}  Theoretical predictions of the CGC saturation model~\cite{Mantysaari:2019jhh} (left) and the LTA nuclear shadowing model~\cite{Guzey:2021star} (right). Coherent and incoherent contributions from the two models are presented separately by dashed lines.  }
\end{figure*}

To further understand the structure of gluons in the deuteron, we compare our data quantitatively with aforementioned  theoretical models - CGC~\cite{Mantysaari:2019jhh} and LTA~\cite{Guzey:2021star}. It is important to note that for STAR kinematics, where Bjorken-$x\sim0.01$, a very small gluon saturation or the nuclear shadowing effect is expected. Without these effects, however, the data and model comparisons (and comparison between models) will be more sensitive to the underlying gluon density distributions, deuteron breakup processes, etc. There are a few model variations available for comparison with the STAR data, while only one variation from each model is presented in Fig.~\ref{fig:figure_3}. The presented CGC and LTA predictions use the AV18 deuteron wavefunction~\cite{Wiringa:1994wb} with effects of nucleon shape and cross section fluctuations, respectively~\cite{Mantysaari:2019jhh, Guzey:2021star}. Other model variations and their comparisons to the data are available in the Supplementary Material, which includes Refs.~\cite{Mantysaari:2019jhh,Miller:2007ri,Wiringa:1994wb,Guzey:2021star}. In Fig.~\ref{fig:figure_3}, the sum of all diffractive processes (coherent and incoherent) are presented for both models, and denoted by lines. The ratios between the total data and the two models are shown in the lower panel. Note that the theoretical uncertainties related to these two models are significantly less than those of the data in the measured $-t$ range, and therefore are not shown. 

It is found that the prediction based on the CGC model describes the data better quantitatively, where the $\chi^{2}$ per degree of freedom is found to be 3.38. On the other hand, the LTA overpredicts the data over most of the measured $-t$ range except for the first bin, resulting in a $\chi^{2}$ per degree of freedom of 13.41. In these analyses, no model parameters are allowed to vary, thus the absolute differential cross sections from the models are directly compared with the data. Although the small number of degrees of freedom might make the absolute $\chi^{2}$ values suspect, their relative sizes for the two models is still highly relevant.

In Fig.~\ref{fig:figure_4}, our total and $n$-tagged data are compared with the same model predictions from Fig.~\ref{fig:figure_3}, but decomposed into coherent and incoherent contributions. For the coherent process, the LTA predicts a similar $-t$ distribution as that of the CGC, where the slope of the coherent $-t$ distribution is generally a measure of the target size~\cite{Toll:2012mb}. In contrast, the incoherent contributions are found to be significantly different, especially at low $-t$, which is in a regime that is sensitive to the deuteron breakup. No experimental data were available in this kinematic region prior to this measurement. Therefore, by using the forward neutron tagging in the ZDC, the $n$-tagged data in this Letter provide the first direct measurement of incoherent diffractive \jpsi production at low $-t$. The result is found to be in better agreement with the incoherent prediction based on the CGC model. A quantitative comparison between the $n$-tagged data and incoherent contributions from the two models can be found in the Supplementary Material.  

In conclusion, the differential cross section of \jpsi photoproduction has been measured as a function of momentum transfer $-t$ in \dAu ultra-peripheral collisions at $\sqrt{s_{_{\rm NN}}}=200~\rm GeV$ using the STAR detector. The data are corrected to the photon-deuteron center-of-mass system, where all final-state particles from deuteron breakup are included. In addition, the differential cross section with a single neutron detected in the deuteron-going Zero-Degree Calorimeter is reported. The data are compared with theoretical predictions based on the Color Glass Condensate saturation model and the Leading Twist Approximation nuclear shadowing model. Both models use the same paradigm\footnote{The Good-Walker paradigm: the coherent production probes the average nuclear distribution, while the incoherent production is sensitive to event-by-event changes in the nuclear configuration~\cite{PhysRev.120.1857}. } to describe the coherent and incoherent photoproduction of $J/\psi$ in ultra-peripheral collisions. The saturation model approaches the problem with dynamical modeling of the gluon density and its fluctuation of the target, while the nuclear shadowing model emphasizes the importance of a shadowing correction from multi-nucleon interaction in nuclei and the fluctuation of the dipole cross section.  The data are found to be in better agreement with the saturation model for incoherent production, where the disagreement between the two models has provided important insights into our theoretical understanding of the nuclear breakup processes. 

Understanding these processes in a simple nuclear environment will be indispensable to further understanding the nuclear effect in heavy nuclei. The data and model comparisons reported in this Letter place significant experimental constraints on the deuteron gluon density distributions and the deuteron breakup process. The results reported here of \jpsi photoproduction will serve as an essential experimental baseline for a high precision measurement of diffractive \jpsi production at the upcoming Electron-Ion Collider.  

We thank the RHIC Operations Group and RCF at BNL, the NERSC Center at LBNL, and the Open Science Grid consortium for providing resources and support.  This work was supported in part by the Office of Nuclear Physics within the U.S. DOE Office of Science, the U.S. National Science Foundation, the Ministry of Education and Science of the Russian Federation, National Natural Science Foundation of China, Chinese Academy of Science, the Ministry of Science and Technology of China and the Chinese Ministry of Education, the Higher Education Sprout Project by Ministry of Education at NCKU, the National Research Foundation of Korea, Czech Science Foundation and Ministry of Education, Youth and Sports of the Czech Republic, Hungarian National Research, Development and Innovation Office, New National Excellency Programme of the Hungarian Ministry of Human Capacities, Department of Atomic Energy and Department of Science and Technology of the Government of India, the National Science Centre of Poland, the Ministry  of Science, Education and Sports of the Republic of Croatia, RosAtom of Russia and German Bundesministerium f\"ur Bildung, Wissenschaft, Forschung and Technologie (BMBF), Helmholtz Association, Ministry of Education, Culture, Sports, Science, and Technology (MEXT) and Japan Society for the Promotion of Science (JSPS).

\bibliography{reference}

\end{document}